\documentclass[aps,twocolumn,prl,nofootinbib]{revtex4-1}
\usepackage{graphicx}
\usepackage{cancel}
\usepackage{textcomp}
\usepackage{amsmath, amssymb, amsfonts, latexsym, epsfig}
\usepackage{mathrsfs}
\usepackage{bm}
\usepackage{times}
\usepackage{epsfig}
\usepackage{color}

\begin{document}

\title{Light Top Squark in Precision Top Quark Sample}

\author{Xue-Qian Li~$^{a}$}

\author{Zong-Guo Si~$^{b}$}

\author{Kai Wang~$^{c}$, \linebreak Liucheng Wang~$^{c,d}$}

\email{lcwang@udel.edu}
\thanks{(Corresponding Author)}

\author{Liangliang Zhang~$^{a,c}$}

\author{Guohuai Zhu~$^{c}$}

\affiliation{$^{a}$~School of Physics, Nankai University, Tianjin 300071, China\\
 $^{b}$~Department of Physics, Shandong University, Jinan, Shandong
250100, China\\
 $^{c}$~Zhejiang Institute of Modern Physics, Department of Physics, Zhejiang University, Hangzhou, Zhejiang 310027, China\\
$^{d}$~Bartol Research Institute, Department of Physics and Astronomy, University of Delaware, Newark, DE 19716, USA
}

\begin{abstract}
The uncertainty of  $t\bar{t}$ production cross section measurement at LHC is at a-few-percent level which 
still allows the stop pair production $\tilde{t}\tilde{t}^{*}$ with identical final states $2b+\ell+nj+\cancel{E}_{T}$. 
In this paper, we attempt to use the existing measurement of $W$-polarization in top quark decay 
to improve the distinction between stop and top quark states. We apply the ATLAS $\chi^2$ method of
$W$-polarization measurement in semi-leptonic $t\bar{t}$ final state to semi-leptonic stop pair samples and study its
prediction. We find that the faked top events from stop mostly contribute to the left-handed polarized 
$W$ due to the reconstruction. The benchmark point with maximal contribution 
to top events only changes $F_{L}$ by 1\%.  After comparing with the current experiments, we conclude that 
the current measurement of $W$-polarization in $t$ decay cannot exclude the light stop scenario. 
\end{abstract}

\maketitle

\section{Introduction}
A Higgs boson with mass around 125~GeV has been discovered at the Large Hadron Collider (LHC) \cite{Aad:2012tfa, Chatrchyan:2012ufa}. If the Higgs boson is a fundamental scalar, quadratic divergence of the quantum correction to its mass is a major concern from the theoretical perspective. Low energy supersymmetry (SUSY) provides an elegant solution to this problem and would be a natural extension of the Standard Model (SM). In SUSY models, the top quark correction to the Higgs mass would be canceled by the contribution from top squark (stop). Stop have been directly searched at the LHC via various channels but no excess was observed. One may wonder if it really exists, but hides itself in some processes or has been misidentified \cite{Bai:2013ema,Cao:2012rz,Han:2013kga,Han:2013usa,Evans:2013jna,1402,Buckley:2014fqa}. 
Actually, exclusion bound of any signal at this QCD machine is model dependent. Scenarios with compressed spectrum always suffer from huge irreducible background. For stop search, its exclusion limit depends on the assumptions about the mass hierarchy of stop, charginos and neutralinos \cite{ATLAS-CONF-2013-037,ATLAS-CONF-2013-048}. For the lightest chargino $\tilde{\chi}_{1}^{\pm}$ and neutralino $\tilde{\chi}_{1}^{0}$, the latest bound via direct search of tri-lepton plus $\cancel{E}_{T}$ indeed leaves a corner of $M_{\tilde{\chi}_{1}^{\pm}}\simeq150$~GeV with $M_{\tilde{\chi}_{1}^{0}}\simeq100$~GeV \cite{ATLAS-CONF-2013-035} \footnote{Search of charginos from squark or gluino cascade decay in jets with lepton $nj+\cancel{E}_{T}+\ell^{\pm}$
puts stronger bounds on chargino masses but it's also model-dependent so we didn't take it account. Instead, we only take the direct search bounds which is more independent of model assumptions. }. On such a mass assumption, stop with mass around 190~GeV decaying with 100\% branching ratio (BR) to bottom quark and $\tilde{\chi}_{1}^{\pm}$ survives existing tests \cite{ATLAS-CONF-2013-048}, because it is difficult to distinguish the stop events from the enormous top background with identical final states. At the LHC with central energy 8~TeV, the $t\bar{t}$ cross section is \cite{Collaboration:2013bma}
\begin{equation}
\sigma_{t\bar{t}}^{\sqrt{s}=8~{\rm TeV}}=241\pm2({\rm stat.})\pm31({\rm syst.)\pm9({\rm lumi.})~\text{pb}},
\end{equation}
which is consistent with the theoretical prediction $\sigma_{t\bar{t}}^{\mathrm{th}}=238_{-24}^{+22}~\text{pb}$.
For 200~GeV stop, its pair production is only 6~pb. Even for 180~GeV stop, its pair production
is around 20~pb, which is still within the error bar of $t\bar{t}$ events.

On the other hand, the prompt decay of $t$ before hadronization provides an opportunity to make its precision measurement possible. The $W$ boson from $t$ decay can be produced either left-handed, right-handed or longitudinal, with the helicity fractions either $F_{L}$, $F_{R}$ or $F_{0}$ respectively. The exact values of helicity fractions can be determined experimentally by detecting the moving direction of the lepton from $W$-boson decay \cite{Kane:1991bg}. Because of the angular momentum conservation and neutrino being only left-handed, the normalized decay rate of $t\rightarrow W^+b,W^+\rightarrow l^+\nu$ is \cite{Kane:1991bg,Hubaut:2005er}
\begin{equation}
\frac{1}{\Gamma}\frac{d\Gamma}{d\cos\theta}=\frac{3}{8}F_{L}(1-\cos\theta)^{2}+\frac{3}{8}F_{R}(1+\cos\theta)^{2}+\frac{3}{4}F_{0}\sin^{2}\theta.\label{eq:theta}
\end{equation}
Here $\theta$ is defined as the angle between the momentum direction of the lepton from the $W$-boson decay and the reversed momentum direction of the bottom quark from top-quark decay, 
both boosted into the $W$-boson rest frame. Precision studies of $W$-boson polarization from top-quark decay have already been performed in the 7~TeV-8~TeV running of the LHC \cite{Aad:2012ky,ATLAS-CONF-2011-037,ATLAS-CONF-2013-033,CMS-PAS-TOP-11-020}. Measurements of these helicity fractions are very important to test the $V$-$A$ structure of the SM, the Higgs mechanism, and the perturbative QCD (pQCD) calculation. Including the finite bottom-quark mass and electroweak effects,  next-to-next-to-leading
order (NNLO) pQCD predictions are $F_{L}=0.311$, $F_{R}=0.0017$ and $F_{0}=0.687$ \cite{Czarnecki:2010gb}.

If events of light stop pairs exist, they can also contribute to those precision measurements. In this paper, we attempt to use the existing measurement of $W$-polarization in top sample to distinguish the light stop events, if they do exist, from the SM $t\bar{t}$ background. Experimentally, such a precision measurement requires a full reconstruction of $W$ boson and top quark. In our studies, we employ the $\chi^{2}$ method given by the ATLAS Collaboration \cite{ATLAS-CONF-2011-037} to reconstruct events. We require the final states of semileptonic events to contain an isolated lepton (electron or muon), missing transverse momentum $\cancel{E}_{T}$ and four jets with following $p_{T}$ requirements \cite{ATLAS-CONF-2011-037}: \begin{table}[tbh]
\begin{center}
\begin{tabular}{|c|c|c|c|c|c|c|}
\hline
& $M_{\tilde{t}},M_{\tilde{\chi}^{\pm}_{1}},M_{\tilde{\chi}^{0}_1}$~(GeV) & $\sigma_{\tilde{t}\tilde{t}^{*}}\text{BR}$~(pb) & $\epsilon$ & $\epsilon K \sigma_{\tilde{t}\tilde{t}^{*}} \text{BR}$~(pb)\\
\hline
\hline
A & 150,~110,~80 & 8.205 & 0.06\% & 0.082\\
\hline
B & 160,~115,~85 &  5.758 & 0.82\% & 0.08\\
\hline
C &170,~130,~90 & 4.222 & 1.35\% & 0.097\\
\hline
D &180,~130,~90 & 3.067 & 1.76\% & 0.092\\
\hline
E &190,~140,~95 & 2.27 & 2.49\% & 0.096\\
\hline
F & 200,~150,~100 & 1.742 & 3.13\% & 0.0927\\
\hline
\end{tabular}
\end{center}
\caption{Survival probabilities after cuts for stop events are shown for LHC@8TeV. NLO QCD $K$-factor is taken to be 1.7.}
\label{cut}
\end{table}
\begin{itemize}
\item $p_{T}>20$ GeV for an isolated electron or muon;
\item $p_{T}>25$ GeV and $\left|\eta\right|<2.5$ for a jet.
\end{itemize}
Stop pair events and $t\bar{t}$ pairs have very different
kinematic features and the events probabilities that pass the selection cuts
are also different. We use the survival probability after cuts $\epsilon$ to quantize the bounds. The $p_{T}$ of $b$-jet or lepton significantly depends on the mass differences $M_{\tilde{t}}-M_{\tilde{\chi}^{\pm}_{1}}$ or $M_{\tilde{\chi}^{\pm}_{1}}-M_{\tilde{\chi}^{0}_{1}}$. We assume the sleptons are much heavier than chargino to minimize the flavor violation. In Table \ref{cut}, we list survival probabilities
of a few benchmark points. We also simulate the survival probability for semi-leptonic $t\bar{t}$ in SM at 8~TeV LHC as $\epsilon^{\rm SM}_{t\bar{t}}= 14.42$.
The benchmark point $C$ with maximal final rate in Table \ref{cut} only corresponds to the effective $t\bar{t}$ cross section  as
\begin{equation}
0.097/\epsilon^{\rm SM}_{t\bar{t}} / \text{BR} = 2.33~\text{pb},
\end{equation}
which is within the uncertainty of cross section measurement.

This paper is organized as follow. Section II is devoted to the polarization in light stop scenario. We show our results after reconstruction in Section III and conclude in section IV.

\section{Polarization in light stop scenario}

In the decoupling MSSM limit, the 125~GeV new particle can be identified
as the lightest CP-even Higgs boson $h$. In order to
predict $m_{h}$=125~GeV, a large $A_{t}$ is necessary
for moderate light stops \cite{Hall:2011aa,Draper:2011aa,Carena:2011aa,Arbey:2011ab,Heinemeyer:2011aa,Ke:2012qc,Ke:2012zq,Ke:2012yc,Han:2013mga}.
Such a large $A_{t}$-term would lead to a big splitting of stop masses.
So the scenario with a light $\tilde{t}_{1}$ is possible and $\tilde{t}_{1}$ would be $50\%$ of $\tilde{t}_{L}$ and $50\%$ of $\tilde{t}_{R}$. These features may be changed in NMSSM. Due to the extra contribution coming from the singlet, it would be easier to realize a 125~GeV Higgs boson in NMSSM compared to MSSM \cite{Cao:2012fz,Cao:2013gba}. A large $A_{t}$-term may not be necessary. So the mass eigenstate $\tilde{t}_{1}$ may be pure $\tilde{t}_{L}$ or pure $\tilde{t}_{R}$ in NMSSM.

The relevant Lagrangians involving $\tilde{t}_{1}\rightarrow b\tilde{\chi}_{1}^{+}$ are \cite{Low:2013aza,Haber:1984rc}
\begin{equation}
\mathcal{L}=\{[-gV_{11}\tilde{t}_{L}+y_{t}V_{12}\tilde{t}_{R}]\bar{b}P_{R}+y_{b}U_{12}\tilde{t}_{L}\bar{b}P_{L}\}\tilde{\chi}_{1}^{+c}.\label{eq:lagrangian}
\end{equation}
Here $U_{ij}$ ($V_{ij}$) are the neutralino (chargino) mixing matrices
and $y_{t}$ ($y_{b}$) is the top (bottom) Yukawa coupling. The mass eigenstate $\tilde{\chi}_{1}^{+}$ is combined by wino and higgsino. If $\tilde{\chi}_{1}^{\pm}$ is pure wino-like, the $\tilde{t}_{L}$ component of $\tilde{t}_{1}$ decays into $\tilde{\chi}_{1}^{+}$ via the first term of Eq.(\ref{eq:lagrangian}). $\tilde{\chi}_{1}^{+}$ is always left-handed. If $\tilde{\chi}_{1}^{\pm}$ is pure higgsino-like, $\tilde{\chi}_{1}^{+}$ from stop decay can be left-handed via the second term of Eq.(\ref{eq:lagrangian}) and right-handed via the third term. The right-handed helicity of $\tilde{\chi}_{1}^{+}$ can be significant at a large $\tan\beta$, since $y_{b}$ can be significantly enhanced at a large $\tan\beta$ compared to $y_{t}$, as $\frac{y_{b}}{y_{t}}=\frac{m_{b}}{\sqrt{2}v_{d}}/\frac{m_{t}}{\sqrt{2}v_{u}}=\frac{m_{b}}{m_{t}}\tan\beta$.
\begin{figure}
\includegraphics[scale=0.215]{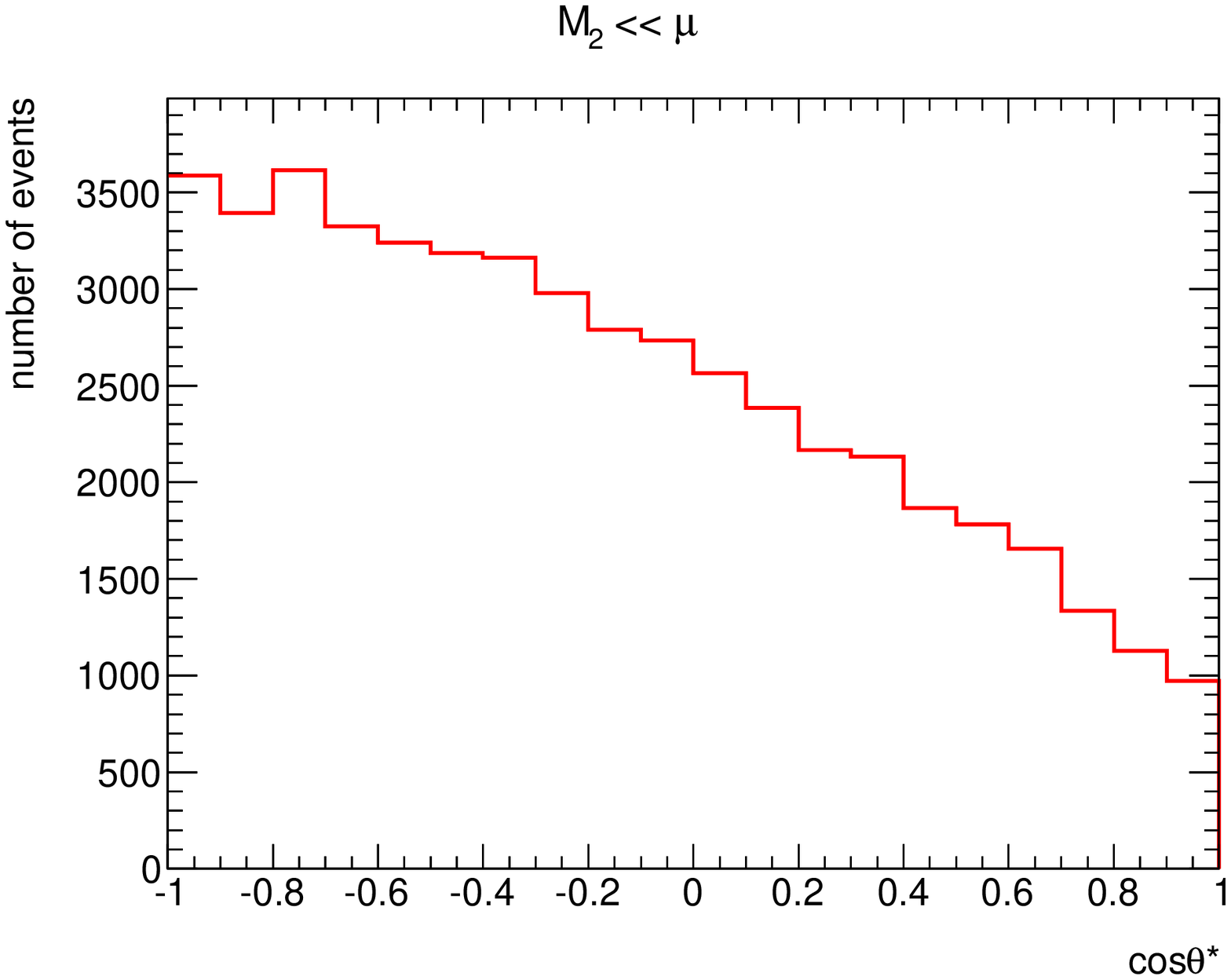}\includegraphics[scale=0.215]{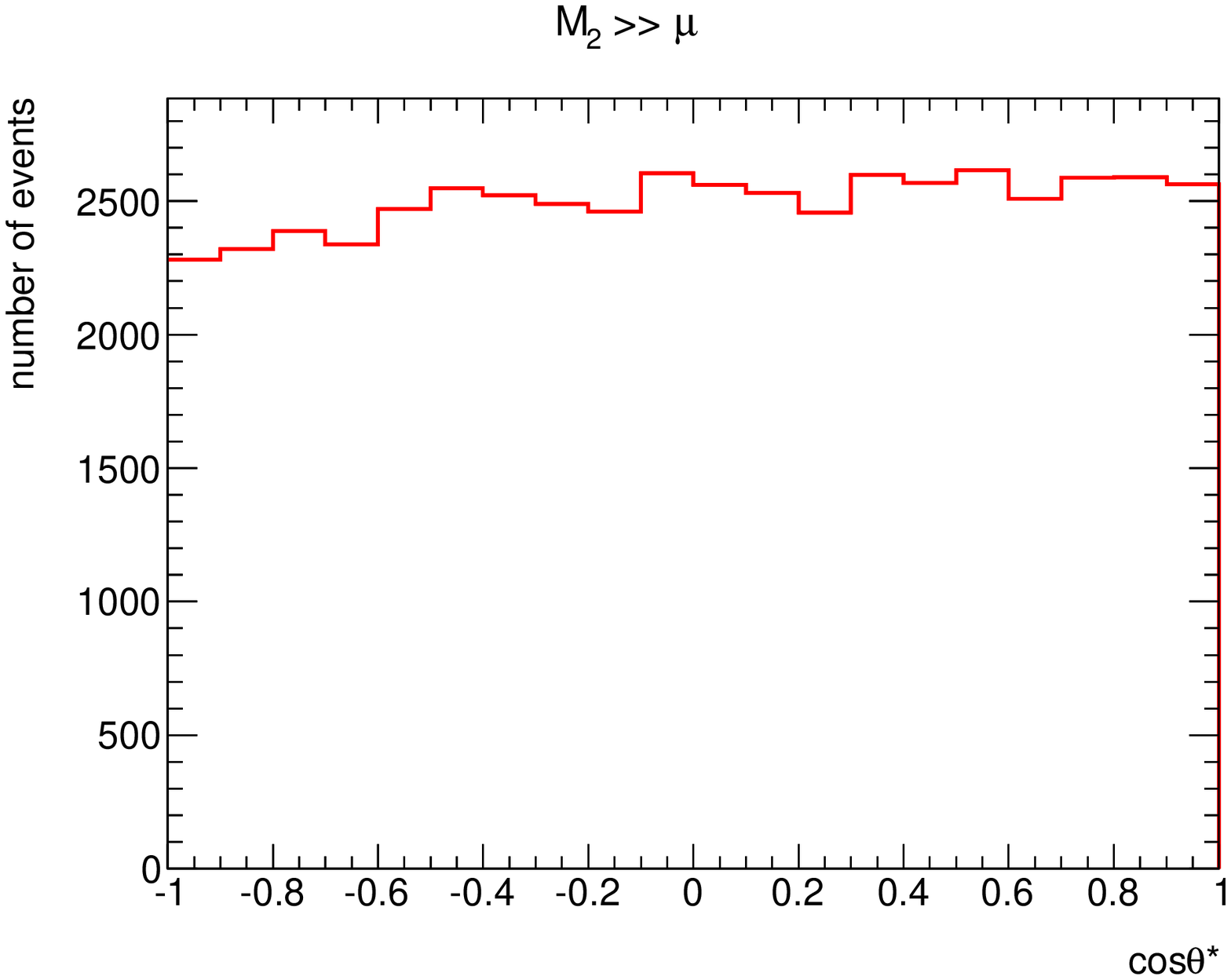}\caption{\label{fig:1} The $\cos\theta$ distribution in the $\tilde{\chi}_{1}^{+}$ rest frame at parton level. Left: If $M_{2}\ll\mu$, $\tilde{\chi}_{1}^{+}$ is mainly left-handed.
Right: If $\mu\ll M_{2}$ with a large $\tan\beta$, the right-handed helicity of $\tilde{\chi}_{1}^{+}$ can be significant.}
\end{figure}

Similar to $t\rightarrow W^+b,W^+\rightarrow l^+\nu$, there exists some spin correlation for the cascade decay $\tilde{t}_{1}\rightarrow b\tilde{\chi}_{1}^{+},\tilde{\chi}_{1}^{+}\rightarrow\tilde{\chi}_{1}^{0}l^+\nu$ \cite{Wang:2013nwm}. The angular distribution of the lepton from chargino decay can be used to partially probe the chargino polarization. We define the angle $\theta$ between the momentum direction of the lepton from chargino decay and the reversed momentum direction of bottom quark from stop decay, both boosted into the chargino rest frame. Because of angular momentum conservation, the distribution of $\cos\theta$ would in general peak at $\theta=\pi~(0)$ for left (right)-handed $\tilde{\chi}_{1}^{+}$ \cite{Wang:2013nwm}. To illustrate this feature, we consider two extreme cases in this paper. One is under the assumption $M_2\ll\mu$, which leads to a nearly wino-like $\tilde{\chi}_{1}^{+}$. And $\tilde{\chi}_{1}^{+}$ from stop decay is always left-handed. The other case bases on the assumption $M_2\gg\mu$ as well as a large $\tan\beta$, where the right-handed helicity of $\tilde{\chi}_{1}^{+}$ can be significant. In both cases we assume $m_{\tilde{t}_{1}}=200$~GeV, $\tilde{\chi}_{1}^{\pm}=150$~GeV, $\tilde{\chi}_{1}^{0}=100$~GeV and all masses of sleptons around 1~TeV. We use the code \textsf{Madgraph5} \cite{Alwall:2011uj} to simulate the whole process. The $\cos\theta$ distribution can be exactly obtained in the $\tilde{\chi}_{1}^{+}$ rest frame at parton level, as shown in Fig.(\ref{fig:1}). The right-handed helicity of $\tilde{\chi}_{1}^{+}$ can be significant if $\tilde{\chi}_{1}^{+}$ is higgsino-like with a large $\tan\beta$. For $W$-boson from top-quark decay, the right-hand helicity fraction $F_R$ is approximately vanishing and severely constrained by experimental data. So in this paper, we would like to study whether such a right-hand helicity of $\tilde{\chi}_{1}^{+}$ from $\tilde{t}_1$ decay can help to distinguish $\tilde{t}_1$ events from the SM top-quark background.

\section{Results and Analysis}

\begin{figure}
\includegraphics[scale=0.215]{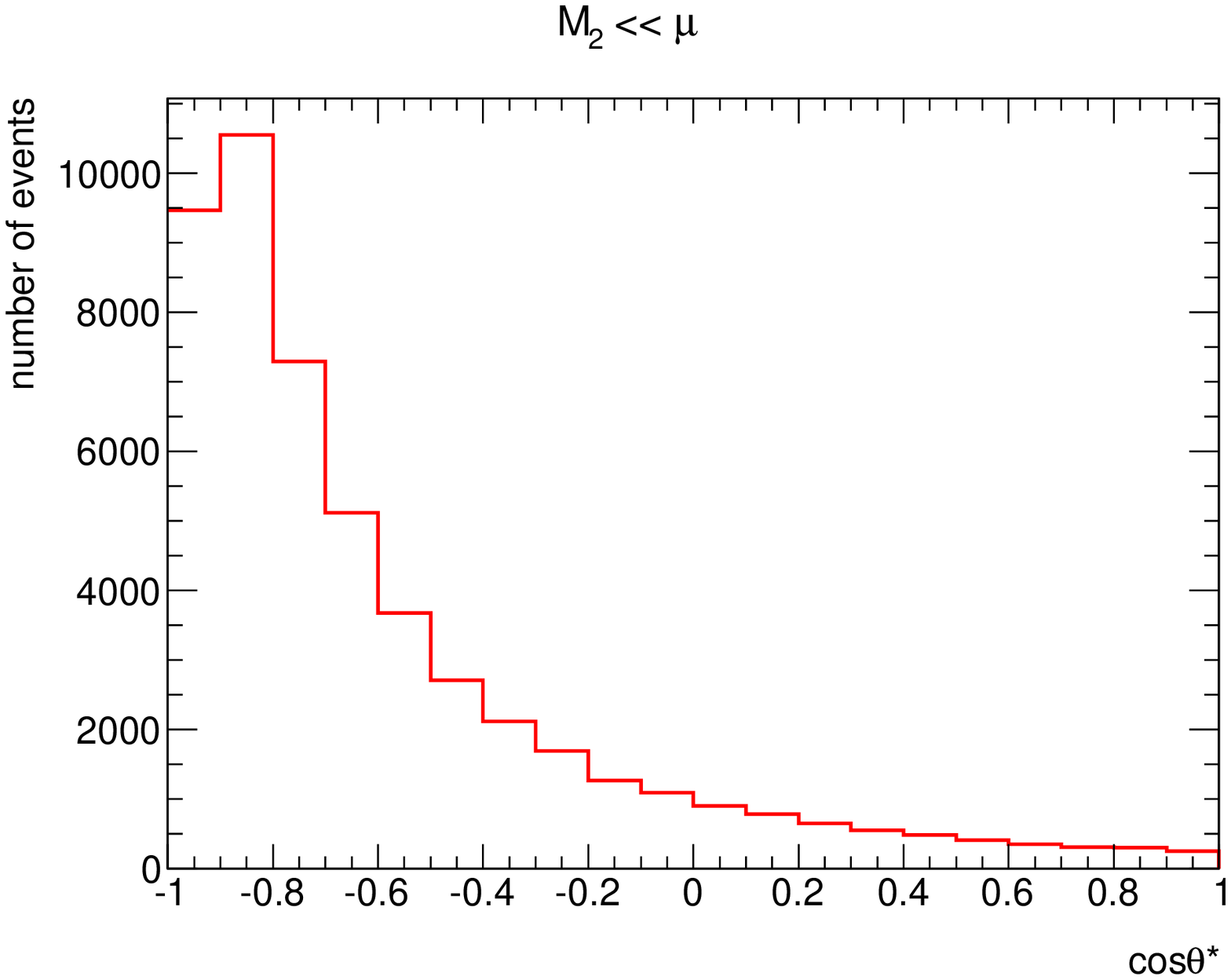}\includegraphics[scale=0.215]{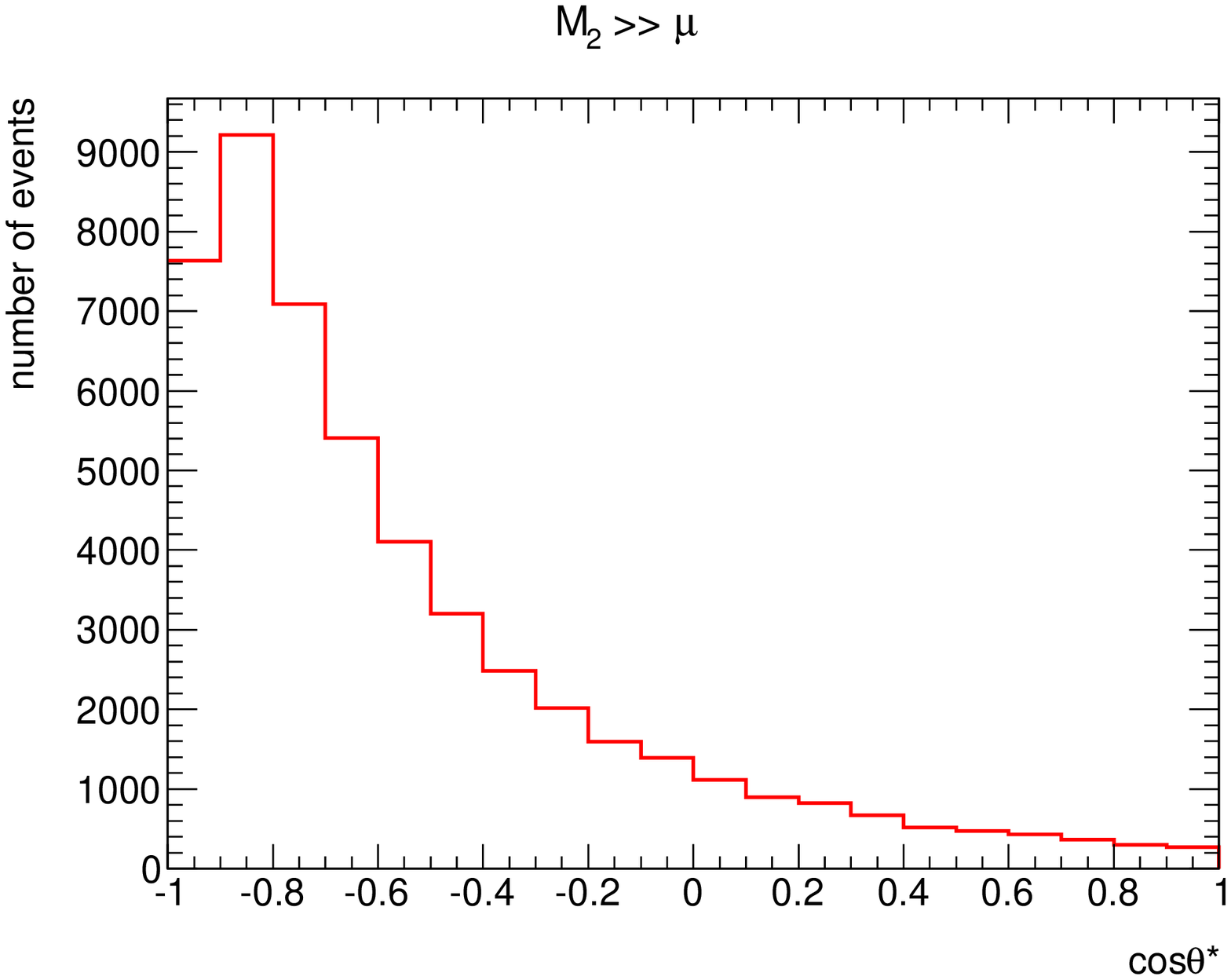}\caption{\label{fig:2} Left: $M_{2}\ll\mu$. Right: $\mu\ll M_{2}$ with a large $\tan\beta$. The $\cos\theta$ distribution of stop events are given in the fake $W$-boson rest frame. Both figures are shown at the parton level. The results after pythia and pgs will be similar to the parton level results.}
\end{figure}

Experimentally, we don't even know there exists chargino from stop
decay. With identical final states, stop pair events will be mis-identified as the SM top events.  The $\cos\theta$ distribution in the $\tilde{\chi}_{1}^{+}$ rest frame, as shown in Fig.(\ref{fig:1}), cannot be experimentally measured. The ATLAS $\chi^2$ method \cite{ATLAS-CONF-2011-037} for $W$-polarization measurement in semi-leptonic $t\bar{t}$ final state will also be applied to semi-leptonic stop pair events. The $\tilde{\chi}_{1}^{0}$ contribution to missing transverse energy $\cancel{E}_{T}$ will always be mis-identified as coming from neutrino. So the $\chi^{2}$ method will always reconstruct a fake neutrino and $W$-boson for each $\tilde{t}_{1}\tilde{t}_{1}^{*}$ event. Moreover, a fake leptonic-branch b-jet may be picked out of the four jets by minimizing the $\chi^{2}$ for some $\tilde{t}_{1}\tilde{t}_{1}^{*}$ events. If so, the reversed momentum of the b-quark is also incorrect. The $\cos\theta$ distribution of stop events can only be obtained in the fake $W$-boson rest frame after $\chi^2$ method, as shown in Fig.(\ref{fig:2}). The left figure is corresponding to $M_{2}\ll\mu$ while the right is corresponding to $\mu\ll M_{2}$ with a large $\tan\beta$. Both $\cos\theta$ distributions peak at $\theta=\pi$, which are generally corresponding to left-handed helicity states.
\begin{figure}
\includegraphics[scale=0.215]{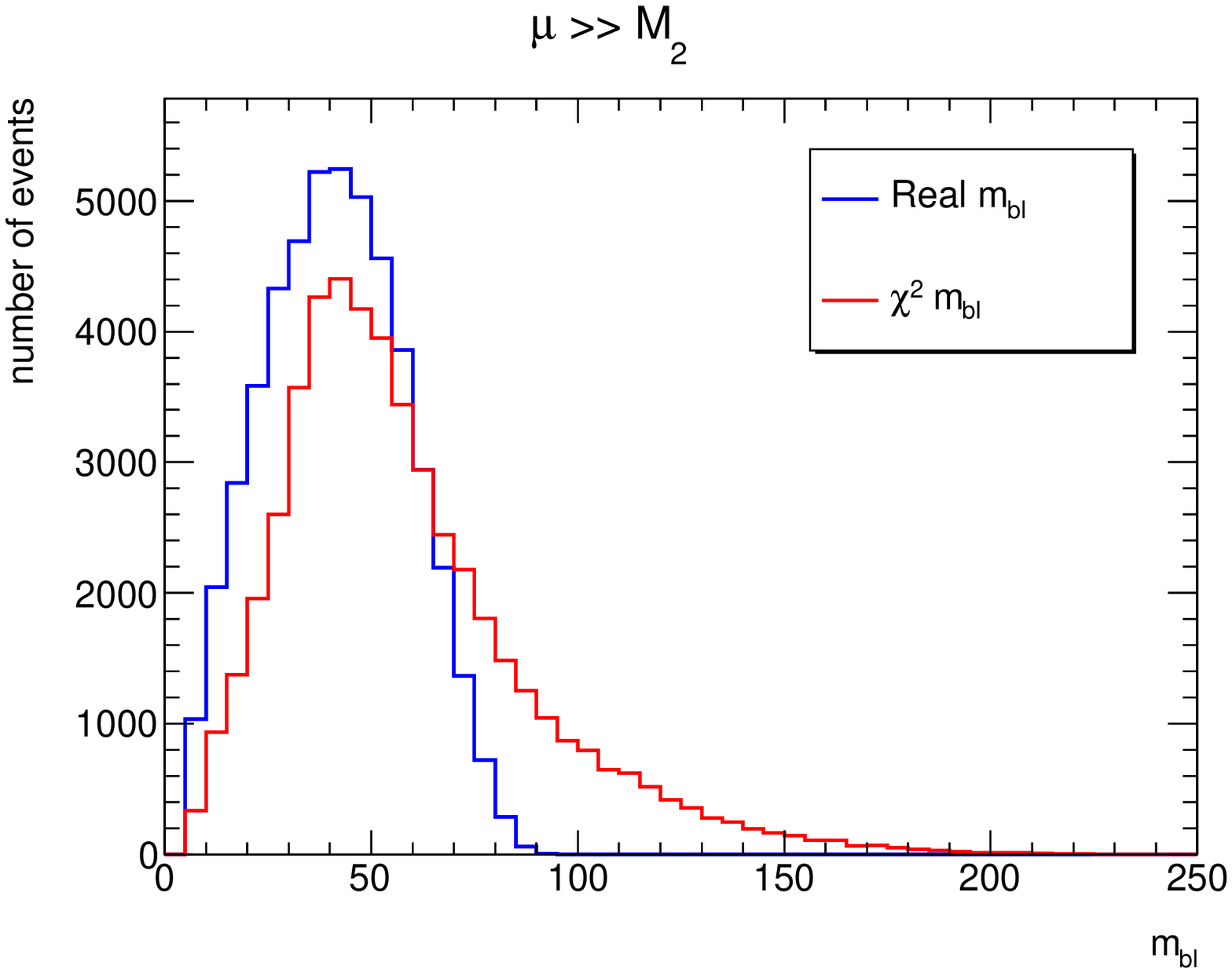}\includegraphics[scale=0.215]{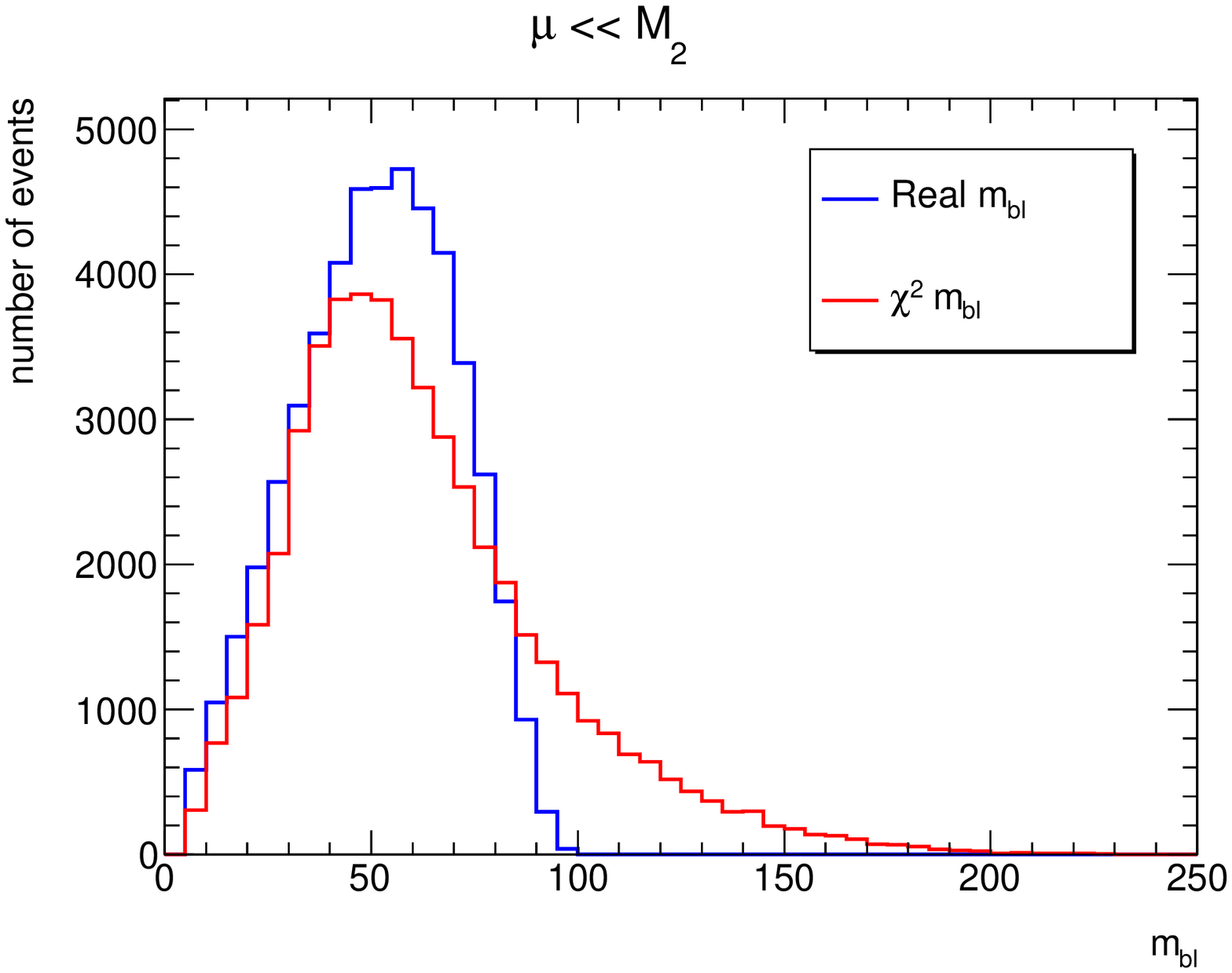}\caption{\label{fig:M_bl}Left: $M_{2}\ll\mu$. Right: $\mu\ll M_{2}$ with a large $\tan\beta$. The blue line is $M_{bl}$ in the real combination
while the red line is fake $M_{bl}$ after $\chi^{2}$ combination, both at the parton level.}
\end{figure}

\begin{figure}
\includegraphics[scale=0.215]{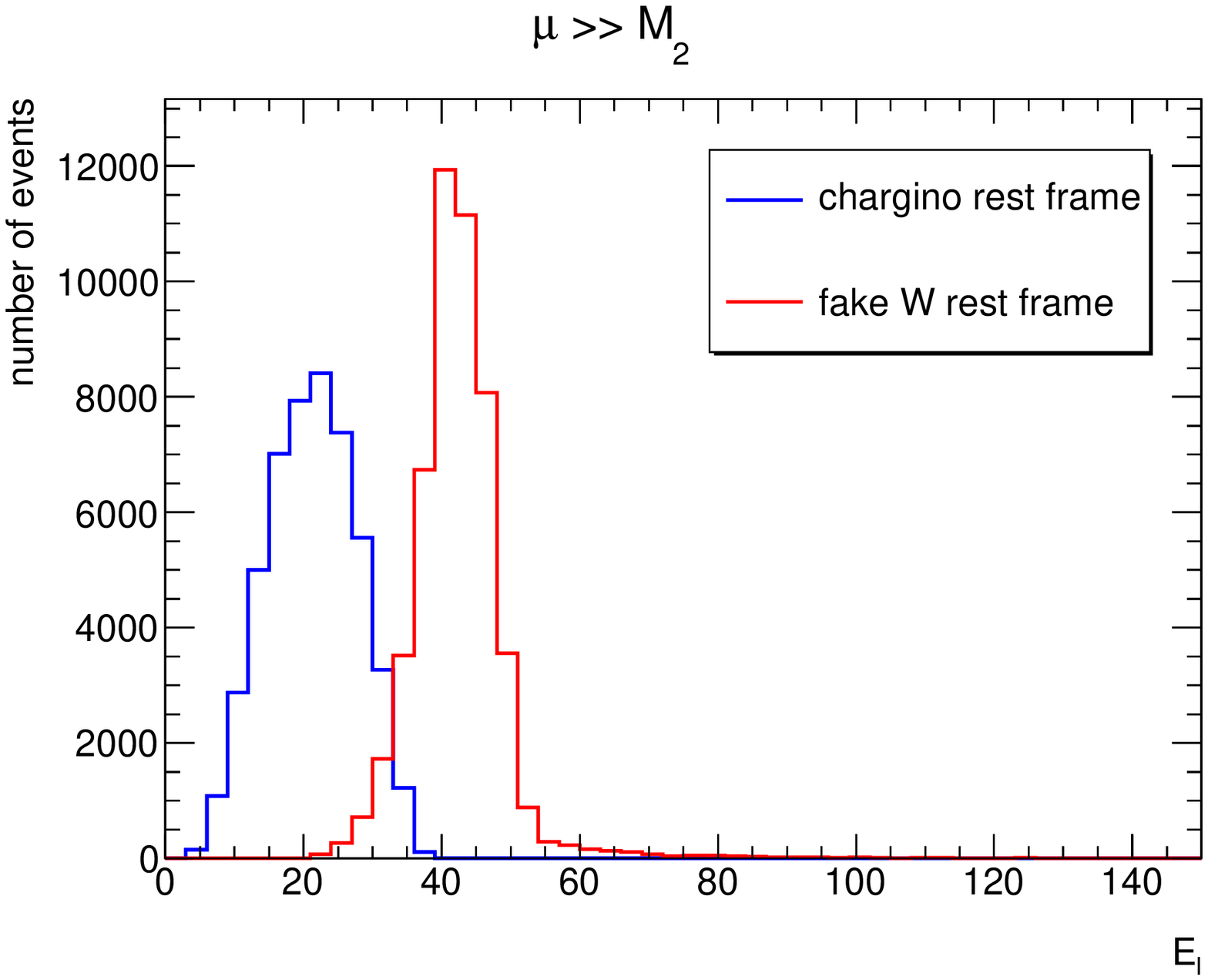}\includegraphics[scale=0.215]{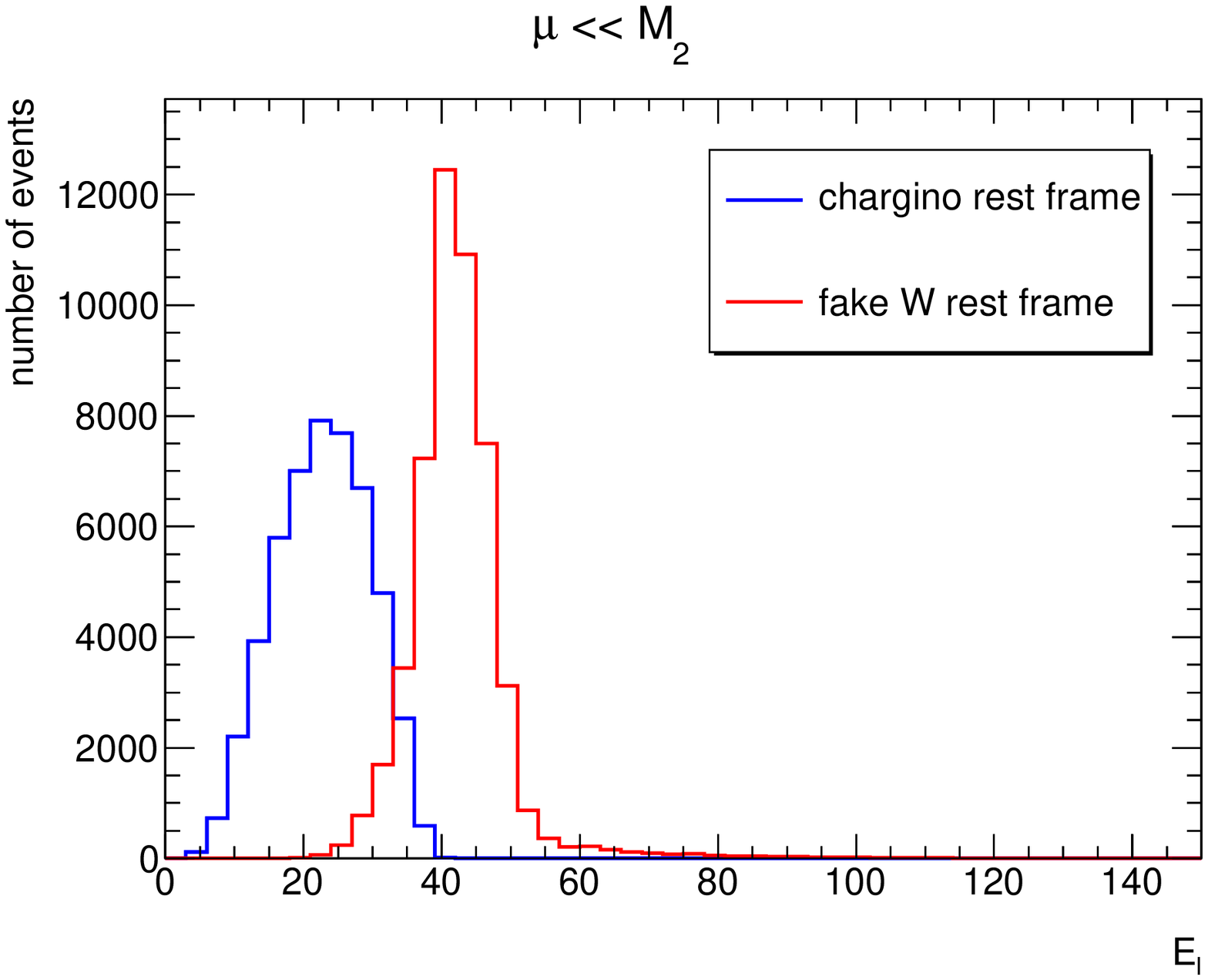}\caption{\label{fig:E_l}Left: $M_{2}\ll\mu$. Right: $\mu\ll M_{2}$ with a large $\tan\beta$. The blue line is $E_{l}$ boosted into the real
$\tilde{\chi}_{1}^{+}$ rest frame while the red line is $E_{l}$
boosted into the fake $W$-boson rest frame after $\chi^{2}$ combination.}
\end{figure}

\begin{figure}
\includegraphics[scale=0.215]{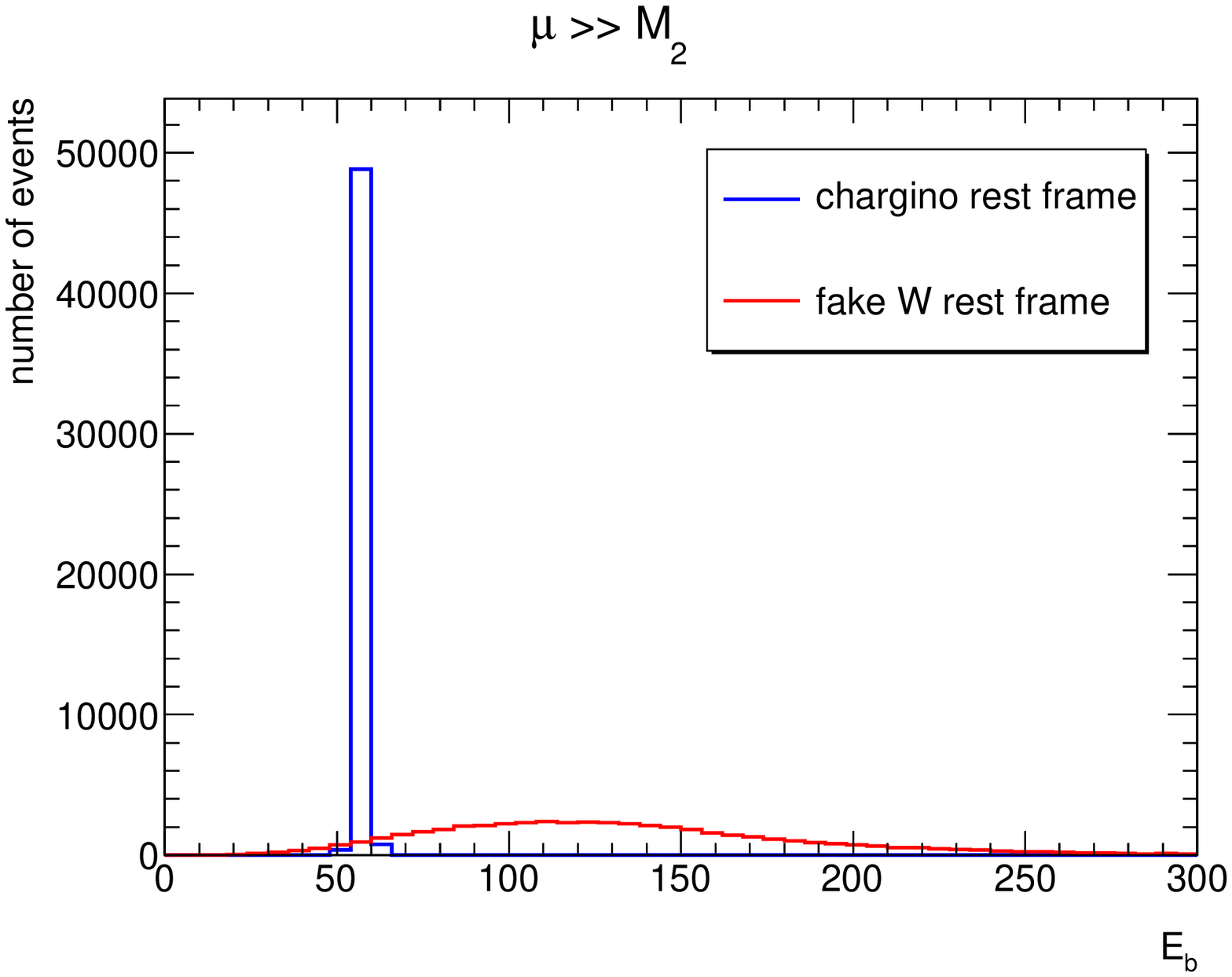}\includegraphics[scale=0.215]{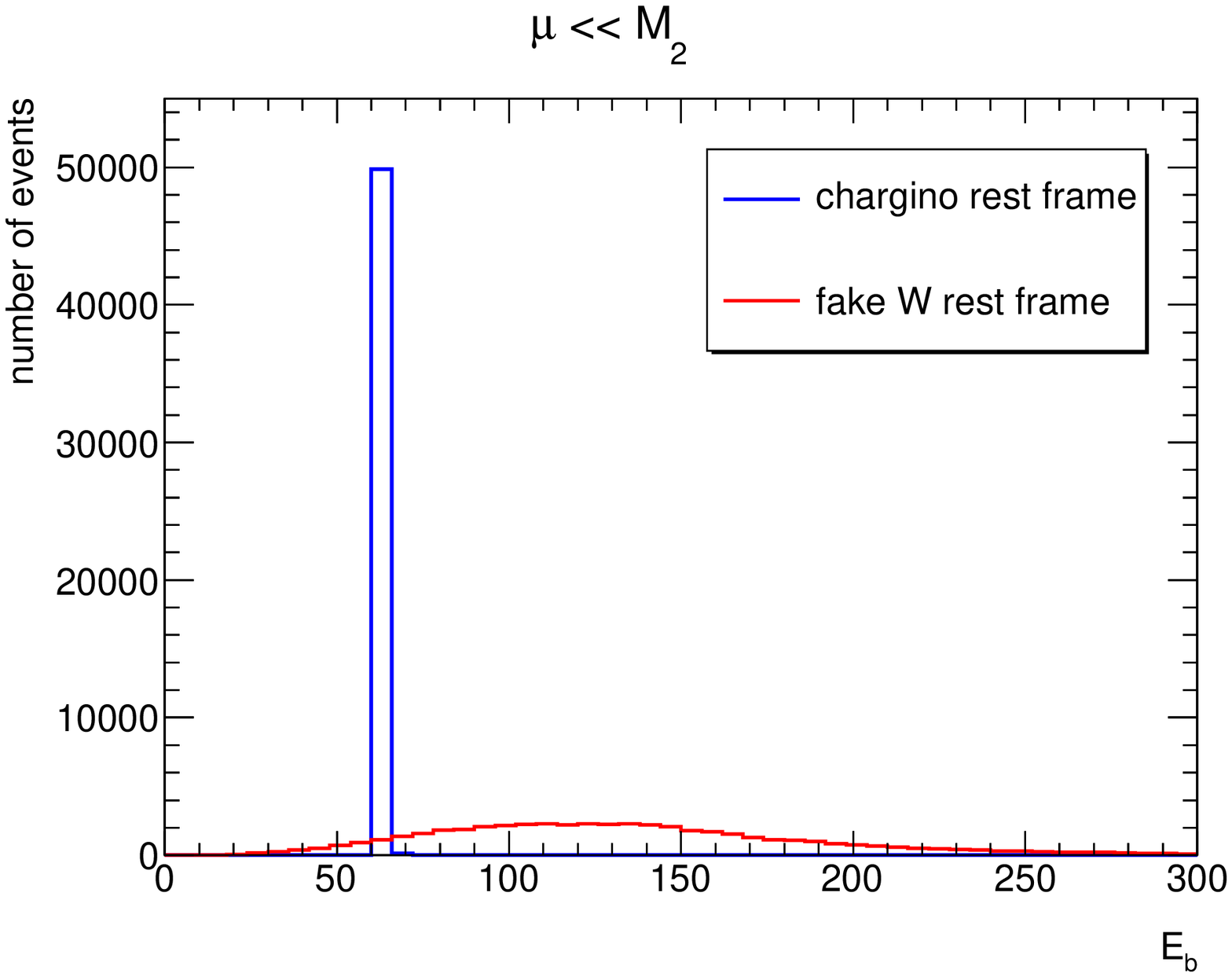}\caption{\label{fig:E_b}Left: $M_{2}\ll\mu$. Right: $\mu\ll M_{2}$ with a large $\tan\beta$. The blue line is $E_{b}$ boosted into the real
$\tilde{\chi}_{1}^{+}$ rest frame while the red line is $E_{b}$
boosted into the fake $W$-boson rest frame after $\chi^{2}$ combination.}
\end{figure}

But we know that the right-handed helicity of the chargino from stop decay is significant under the assumption $\mu\ll M_{2}$ with a large $\tan\beta$, as shown in the right of Fig.(\ref{fig:1}). When the $\chi^2$ method is applied to stop events, the right-handed helicity seems disappeared, as shown in the right of Fig.(\ref{fig:2}). To understand the difference, we focus on a general definition of $\theta$, which is the angle between the momentum direction of the lepton and the reversed momentum direction of the b-quark, both boosted into a specific chosen frame. As the lepton and b-quark are approximately massless,
\begin{equation}
\cos\theta=-\frac{\vec{p}_{l}\cdot\vec{p}_{b}}{\left|\vec{p}_{l}\right|\left|\vec{p}_{b}\right|}=\frac{p_{l}\cdot p_{b}}{\left|\vec{p_{l}}\right|\left|\vec{p_{b}}\right|}-1=\frac{M_{lb}^{2}}{2E_{l}E_{b}}-1\label{eq:new costheta}. 
\end{equation}
Here $M_{lb}$ is the invariant mass of the lepton and b-quark, which is frame-independent. $E_{l}$ ($E_{b}$) is the energy of lepton (b-quark), which
depends on the chosen frame. In order to understand the fake polarization, we show $M_{bl}$,
$E_{l}$ and $E_{b}$ of the stop events in Fig.(\ref{fig:M_bl}), Fig.(\ref{fig:E_l}) and Fig.(\ref{fig:E_b}) respectively.

In Fig.(\ref{fig:M_bl}), the blue line is $M_{bl}$ in the real combination while the red line is $M_{bl}$ after $\chi^{2}$ combination. Since $M_{bl}$ is independent of the chosen frame, the difference between two combinations originates from the wrong combination. By using $\chi^{2}$ method, a fake leptonic-branch b-jet may be picked out of the four jets to get $M_{bl}$. To reduce the wrong combination, we suggest to use b-tagging information in the $\chi^{2}$ reconstruction. However, from Fig.(\ref{fig:M_bl}) we know that two $M_{bl}$ distributions of stop events are similar, which is not the main reason leading to different $\cos\theta$ distributions in Fig.(\ref{fig:1}) and  Fig.(\ref{fig:2}).

In Fig.(\ref{fig:E_l}), $E_{l}$ in the fake $W$-boson rest frame are generally larger than $E_{l}$ in the real $\tilde{\chi}_{1}^{+}$ rest frame. Moreover, the distribution of fake $E_{l}$ peaks at 40~GeV. For $t\bar{t}$ events, we know that energy-momentum conservation requires $E_{l}=M_{W}/2=$40~GeV in the $W$-boson rest frame. So after the $\chi^2$ method is applied to semi-leptonic stop pair events, the $E_{l}$ distribution from stop becomes similar to that from top. In Fig.(\ref{fig:E_b}), $E_{b}$ in the fake $W$-boson rest frame are in general larger than $E_{b}$ in the real $\tilde{\chi}_{1}^{+}$ rest frame. Here $E_{b}$ in the $\tilde{\chi}_{1}^{+}$ rest frame is fixed due to energy-momentum conservation.

\begin{figure}
\includegraphics[scale=0.215]{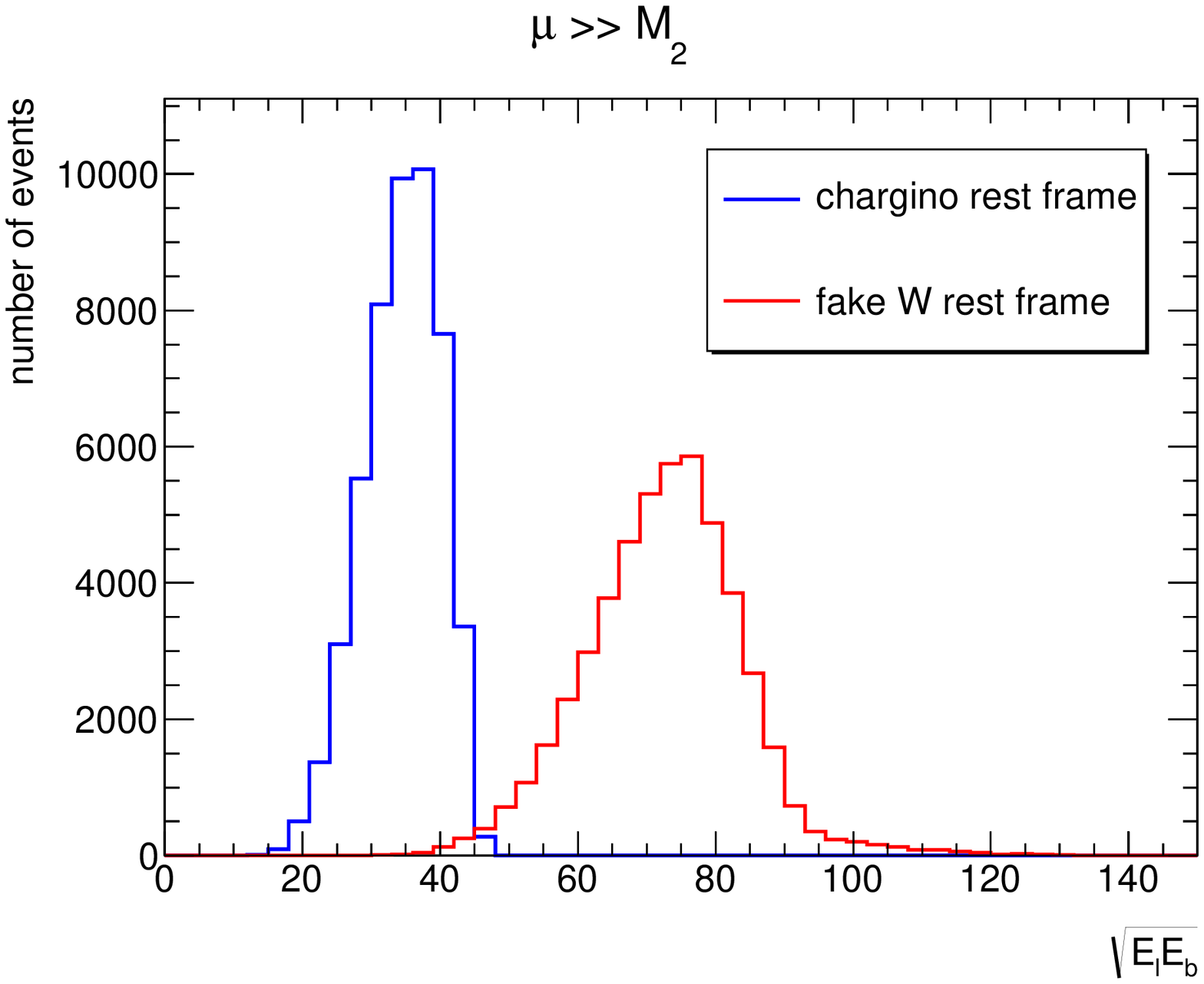}\includegraphics[scale=0.215]{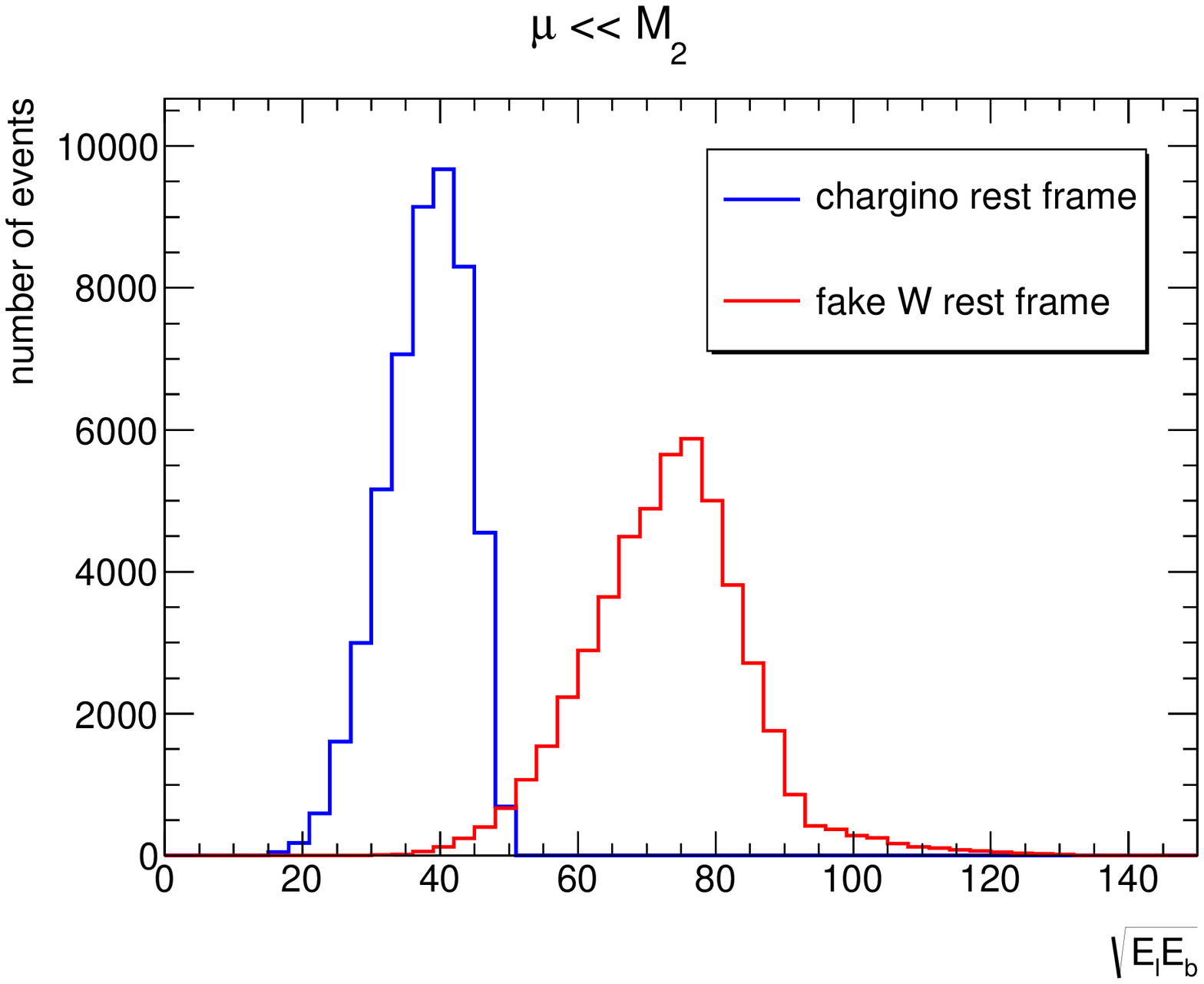}\caption{\label{fig:sqrt of E_bE_l}Left: $M_{2}\ll\mu$. Right: $\mu\ll M_{2}$ with a large $\tan\beta$. The blue line is $\sqrt{E_{b}E_{l}}$ in the real $\tilde{\chi}_{1}^{+}$ rest frame while the
red line is $\sqrt{E_{b}E_{l}}$ in the fake $W$-boson
rest frame after $\chi^{2}$ combination.}
\end{figure}

\begin{figure}
\includegraphics[scale=0.25]{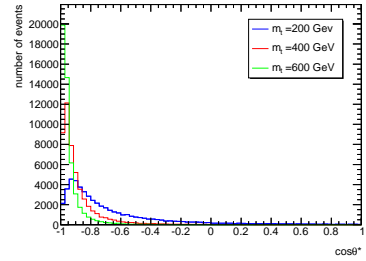}\caption{\label{fig:mt}$\cos\theta$ of stop-pair events are given in the fake $W$-boson rest frame when the top mass $m_{t}$ in the $\chi^{2}$ method are varied.}
\end{figure}

In Fig.(\ref{fig:sqrt of E_bE_l}), we finally show the combination $\sqrt{E_{b}E_{l}}$ of stop-pair events. In both the left and the right figure, $\sqrt{E_{b}E_{l}}$ in the fake $W$-boson rest frame is larger than that in the real $\tilde{\chi}_{1}^{+}$ rest frame. According to Eq.(\ref{eq:new costheta}), the fake $\cos\theta=\frac{M_{lb}^{2}}{2E_{l}E_{b}}-1$ in the $W$-boson rest frame is consequently approaching to -1. That is why the right-handed helicity seems disappeared when the $\chi^2$ method is applied to stop events, 

Actually, when the $\chi^2$ method of semi-leptonic $t\bar{t}$ final state is applied to semi-leptonic stop-pair events, $\sqrt{E_{b}E_{l}}$ of stop event in the fake $W$-boson rest frame
is approximately similar to $\sqrt{E_{b}E_{l}}$ of top event in the $W$-boson rest frame. For stop events after the $\chi^{2}$ method, the distribution in the fake $W$-boson rest frame is approximately governed by the top distribution as
\begin{equation}
\cos\theta=\frac{2M_{bl}^{2}}{m_{t}^{2}-m_{W}^{2}}-1.
\label{eq:top_costheta}
\end{equation}
Here $m_{t}$ and $m_{W}$ are not only physical masses but also expected masses defined in the $\chi^2$ method \cite{ATLAS-CONF-2011-037}.  When the $\chi^2$ method is applied, the final states of stop events are expected to have resonances around $m_{t}$ and $m_{W}$. In order to prove Eq.(\ref{eq:top_costheta}), we vary $m_{t}$ as 200~GeV, 400~GeV and 600~GeV in the $\chi^{2}$ method and the results are shown in Fig.(\ref{fig:mt}). As expected, more stop events fall into the range near $\cos\theta=-1$ when $m_{t}$ becomes larger. The $\cos\theta$ distribution of stop events in the fake $W$-boson rest frame is proved to be governed by Eq.(\ref{eq:top_costheta}).

\section{Conclusion}

A light stop scenario has not yet been excluded, as light stops may be mis-identified as SM top-quarks with identical final states and hide themselves in the uncertainty of the $t\bar{t}$ production cross section. In this paper, we attempt to use the precision measurement of $W$-polarization in top decay to improve the ability to distinguish the light stop from the SM top background. When the ATLAS $\chi^2$ method of semi-leptonic $t\bar{t}$ event is applied to semi-leptonic stop-pair event, the $\cos\theta$ distribution always peaks at $\cos\theta=-1$ due to the fake reconstruction. We compute the $F_{L}/F_{R}/F_{0}$ contribution from benchmark point $C$ in Table.\ref{cut}, which is of the maximal contribution to $t\bar{t}$ events. The result is listed as following:
\begin{eqnarray}
\text{SM} &:& F_{L}=0.303,~F_{R}=0.0493,~F_{0}=0.647,\\
\text{SM+stop} &:& F_{L}=0.313,~F_{R}=0.0497,~F_{0}=0.638.\nonumber
\end{eqnarray}
The benchmark point with maximal contribution to top events only changes $F_{L}$ by 1\%, which is far below the current uncertainty. Thus, we conclude that the current measurement on the $W$-polarization cannot exclude the light stop scenario with stop mass around top-quark mass.

\section*{Acknowledgement}

LW and LZ would like to thank Jiwei Ke for useful discussions about \textsf{Madgraph5}. XL is supported by the National Science Foundation of China (11075079, 11135009). ZS is supported by the National Science Foundation of China (11275114). KW is supported in part, by the Zhejiang University Fundamental Research Funds for the Central Universities (2011QNA3017) and the National Science Foundation of China (11245002,11275168). LW is supported
in part by the DOE Grant No. DE-FG02-12ER41808. GZ is supported in part, by the National Science Foundation of China (11075139, 11135006,11375151) and the Program for New Century Excellent Talents in University (Grant No. NCET-12-0480).

\bibliographystyle{h-physrev}
\bibliography{w_polarization_from_top}

\end{document}